\documentclass[10pt]{article}

\begin{document}

\title{The principle of least action for test particles in a four-dimensional spacetime embedded in $5D$}
\author{J. Ponce de Leon\thanks{E-mail: jpdel@ltp.upr.clu.edu, jpdel1@hotmail.com}\\ Laboratory of Theoretical Physics, Department of Physics\\ 
University of Puerto Rico, P.O. Box 23343, San Juan, \\ PR 00931, USA} 
\date{November  2007}

\maketitle

\begin{abstract}
It is well known that, in the five-dimensional scenario of braneworld and space-time-mass theories,  geodesic motion  in $5D$   is observed to be non-geodesic in $4D$. Usually, the discussion is purely geometric and based on the dimensional reduction of the geodesic equation in $5D$, without any reference to the test particle whatsoever. In this work we obtain the equation of motion in $4D$ directly from the principle of least action. So our main thrust is not the geometry but the particle observed in $4D$.
 A clear physical picture emerges from our work. Specifically, that the deviation from the geodesic motion in $4D$ is due to the variation of the rest mass of a particle, which is induced by the scalar field in the $5D$ metric and the explicit dependence of the spacetime metric on the extra coordinate. 
Thus,  the principle of least action not only leads to the correct  equations of motion,   but also provides a  concrete physical meaning for a number of algebraic quantities  appearing in the geometrical reduction of the geodesic equation.

\end{abstract}

\medskip



\newpage

\section{Introduction}

The possibility that our world may be embedded in a $(4 + d)$-dimensional universe with more than four large dimensions has attracted the attention of a great number of researchers. 
There are several motivations for the introduction of large extra dimensions. 
Among them to resolve the differences between  gravity and quantum field theory; provide possible solutions to the hierarchy and the cosmological constant problems \cite{Arkani1}-\cite{RS2} and ultimately unify all forces of nature.

The idea of large extra dimensions is also inspired by the vision that  matter in $4D$ is purely geometric in nature. In space-time-matter theory (STM) one large extra dimension is needed in order to get a consistent description, at the macroscopic level, of the properties of  the matter as observed in $4D$ \cite{JPdeL 1}-\cite{EMT}. The mathematical support of this theory is given by a theorem of differential geometry due to Campbell and Magaard \cite{Campbell}-\cite{Lidsey}. 

In the Randall $\&$ Sundrum braneworld scenario, STM and other higher-dimensional theories,  the main attempt is to reproduce the physics of four-dimensional gravity up to higher-dimensional modifications to general relativity. From a mathematical viewpoint, this means that the equations in $4D$ are projections of the $5D$ equations on $4D$-hypersurfaces orthogonal to some vector field $\psi^A$, which is identified with the ``extra" dimension. By an appropriate choice of coordinates one can remove the spacetime part of this vector and  put it in the form  
\begin{equation}
\label{e4}
{\psi_A}= (0, 0, 0, 0, {\Phi}).
\end{equation}
In such coordinates the most general line element can be written as\footnote{Lowercase Greek letters go from $0$ to $3$; $x^{0}$ is
 time like, $x^{1}$, $x^{2}$, $x^{3}$ are space like. Capital Latin letters $A, B$ denote indexes in $5D$.} 
\begin{equation}
\label{general metric}
dS^2 = g_{\mu\nu}(x^{\rho}, y)dx^{\mu}dx^{\nu} + \epsilon \Phi^2(x^{\rho}, y)dy^2,
\end{equation}
where $g_{\mu\nu}$ is the metric induced in $4D$; $x^{\mu}$ denote the coordinates of the spacetime; $y$ represents the extra coordinate, and the factor $\epsilon$ can be $- 1$ or $+ 1$ depending on whether the extra dimension is spacelike or timelike, respectively, viz., $\psi_{A}\psi^{A} = \epsilon$.

A possible way of testing for new physics coming from  extra dimensions is to  examine the dynamics of test particles. In practice this means to search for deviations from the universal  ``free fall" in $4D$. The question of how an observer in $4D$, who is confined to making physical measurements in our ordinary spacetime, perceives the motion of test particles governed by the geodesic equation in $5D$
\begin{equation}
\label{geodesic equation in 5D}
\frac{d^2x^{A}}{dS^2} + \Gamma_{BC}^{A}U^{B}U^{C} = 0,\;\;\;U^{A} = \frac{dx^A}{dS},
\end{equation}
has widely been discussed in the literature \cite{MashhoonWesson}-\cite{Romero2}.
The discussion is typically based on the dimensional reduction of geodesics in $5D$, which involves subtle technical details as the choice of adequate affine parameters for the motion in $5D$ and $4D$.
 
After a long and sophisticated calculation,  the dimensional reduction of (\ref{geodesic equation in 5D}) yields
\begin{equation}
\label{Du/ds contravariant}
\frac{d u^{\mu}}{ds} + \Gamma_{\alpha \beta}^{\mu}u^{\alpha}u^{\beta} = \left(\frac{1}{2}u^{\mu}u^{\rho} - g^{\mu\rho}\right)u^{\lambda}\frac{\partial{g_{\rho\lambda}}}{\partial{y}}\left(\frac{dy}{ds}\right) + \epsilon \Phi\left[\Phi^{;\mu} - u^{\mu}u^{\rho}\Phi_{;\rho}\right]\left(\frac{dy}{ds}\right)^2,
\end{equation}
where $\Gamma_{\alpha \beta}^{\mu}$ are the Christoffel symbols calculated with  the spacetime metric $g_{\alpha\beta}$; $ds = \sqrt{g_{\mu\nu}dx^{\mu}dx^{\nu}}$, and $u^{\mu}$ is the usual four-velocity, $u^{\mu} = dx^{\mu}/ds$.
The equation for the covariant components $u_{\mu}$, looks a little simpler, namely,
\begin{equation}
\label{Du/ds covariant}
\frac{d u_{\mu}}{ds} -  \Gamma_{\mu \alpha}^{\beta}u^{\alpha}u_{\beta}  = \frac{1}{2}u_{\mu}u^{\lambda}u^{\rho}\frac{\partial {g_{\lambda\rho}}}{\partial y}\left(\frac{dy}{ds}\right)  + \epsilon \Phi \Phi_{; \rho} \left[\delta_{\mu}^{\rho} - u_{\mu}u^{\rho} \right]\left(\frac{dy}{ds}\right)^2.
\end{equation}

The aim of this paper is twofold. First, to derive these equations from the principle of least action. Second, to give the most general expression for the rest mass of a particle observed in $4D$ in terms of the metric and momentum along the extra dimension.

\section{The principle of least action}

The principle of least action is defined by the statement that for each system there exist an integral $I$, called the action, which has a minimum value for the actual motion, so that its variation $\delta I$ is zero \cite{Landau and Lifshitz}-\cite{comment}. In classical mechanics, the action for a free material point of mass $m$ is the integral $(c = 1)$
\begin{equation}
\label{the action in mechanics}
I_{Mech} = - m\int_{a}^{b}{ds},
\end{equation}
along the world line of the particle between two particular events represented by  $a$ and $b$, for an initial and final position, respectively. 

Among the possible effects of extra dimensions, is the variation of the rest mass $m$. Therefore, we will assume here that the appropriate action is 
\begin{equation}
I = - \int_{a}^{b}{m ds},
\end{equation}
where the {\it function} $m$ is taken at points on the world line of the particle. We now proceed to derive the equations of motion from this action. 

The principle of least action states 
\begin{equation}
\label{The principle of least action states}
\delta I = - \delta \int_{a}^{b}{m ds} = 0. 
\end{equation}
Noting that $m$ as well as the metric are allowed to depend on {\it all} five coordinates, we have 
\begin{equation}
\delta m = \left(\frac{\partial m}{\partial x^{\mu}}\right)\delta x^{\mu} + \left(\frac{\partial m}{\partial y}\right) \delta y,
\end{equation}
and
\begin{equation}
\frac{\delta (ds)}{ds} = \frac{1}{2}u^{\mu}u^{\nu} \frac{\partial g_{\mu\nu}}{\partial x^{\rho}}\delta x^{\rho} + u_{\nu}\frac{d\delta x^{\nu}}{ds} + \frac{1}{2}u^{\mu}u^{\nu} \frac{\partial g_{\mu\nu}}{\partial y}\delta y.
\end{equation}
Substituting these expressions into (\ref{The principle of least action states}) and using that 
\begin{equation}
m u_{\mu}\frac{d\delta x^{\mu}}{ds} = \frac{d}{ds}\left(m u_{\mu} \delta{x^{\mu}}\right) - \left(m \frac{du_{\mu}}{ds} + u_{\mu}\frac{dm}{ds}\right)\delta x^{\mu},
\end{equation}
we obtain
\begin{equation}
\delta S = - \int_{a}^{b}\left\{\left(\frac{1}{2}u^{\mu}u^{\nu}\frac{\partial g_{\mu\nu}}{\partial x^{\rho}} - \frac{du_{\rho}}{ds} - \frac{u_{\rho}}{m}\frac{dm}{ds} + \frac{1}{m}\frac{\partial m}{\partial x^{\rho}}\right)\delta x^{\rho} + \left(\frac{1}{m}\frac{\partial m}{\partial y} + \frac{1}{2}u^{\mu}u^{\nu}\frac{\partial g_{\mu\nu}}{\partial y}\right)\delta y\right\}m ds.
\end{equation}
In integrating by parts, we have used the fact that $\delta x^{\mu} = 0$ at the limits. In view of the arbitrariness of $\delta x^{\mu}$ and $\delta y$, it follows that the integrand is zero, that is 
\begin{equation}
\label{arbitrarity in delta xmu}
\frac{1}{2}u^{\mu}u^{\nu}\frac{\partial g_{\mu\nu}}{\partial x^{\rho}} - \frac{du_{\rho}}{ds} - \frac{u_{\rho}}{m}\frac{dm}{ds} + \frac{1}{m}\frac{\partial m}{\partial x^{\rho}} = 0,
\end{equation}
and 
\begin{equation}
\label{arbitrarity in delta y}
\frac{1}{m}\frac{\partial m}{\partial y} + \frac{1}{2}u^{\mu}u^{\nu}\frac{\partial g_{\mu\nu}}{\partial y} = 0.
\end{equation}
Rearranging terms in (\ref{arbitrarity in delta xmu}) it can be written as
\begin{equation}
\frac{du_{\rho}}{ds} - \Gamma_{\rho \alpha}^{\beta}u^{\alpha}u_{\beta} = \frac{1}{m}\frac{\partial m}{\partial x^{\rho}} - \frac{u_{\rho}}{m}\frac{dm}{ds}.
\end{equation}
From which it follows that 
\begin{equation}
\label{equation of motion in terms of m}
\frac{du_{\rho}}{ds} - \Gamma_{\rho \alpha}^{\beta}u^{\alpha}u_{\beta} = - \frac{u_{\rho}}{m}\frac{\partial m}{\partial y}\frac{dy}{ds} + \frac{1}{m}\frac{\partial m}{\partial x^{\mu}}\left(\delta_{\rho}^{\mu} - u^{\mu}u_{\rho}\right).
\end{equation}
Now using (\ref{arbitrarity in delta y}) we obtain
\begin{equation}
\label{equation of motion with the first term identical to the KK projected}
\frac{du_{\rho}}{ds} - \Gamma_{\rho \alpha}^{\beta}u^{\alpha}u_{\beta} = \frac{1}{2}u_{\rho}u^{\mu}u^{\nu}\frac{\partial g_{\mu\nu}}{\partial y}\left(\frac{dy}{ds}\right) + \frac{1}{m}\frac{\partial m}{\partial x^{\mu}}\left(\delta_{\rho}^{\mu} - u^{\mu}u_{\rho}\right).
\end{equation}
Except for the second term in the r.h.s, this equation is {\it identical} to  (\ref{Du/ds covariant}). Notice that so far we have not specified  the function $m$. 

In order to recover (\ref{Du/ds covariant}), the spacetime derivative of $m$ should satisfy the relation\footnote{To be more precise, since $(\delta^{\mu}_{\nu} - u^{\mu}u_{\nu})$ is the projector onto the tree-space orthogonal to $u^{\mu}$, adding to  (\ref{spatial derivative of m}) an additional function $H u_{\mu}$, with an arbitrary $H$, is innocuous.}
\begin{equation}
\label{spatial derivative of m}
\frac{1}{m}\frac{\partial m}{\partial x^{\mu}} = \epsilon \Phi \frac{\partial \Phi}{\partial x^{\mu}}\left(\frac{dy}{ds}\right)^2.
\end{equation}
For a particle at rest, with respect to the system of coordinates $(dx^{i} = 0)$, the four-velocity becomes
\begin{equation}
\label{particle at rest}
u^{\mu} = \frac{\delta^{\mu}_{0}}{\sqrt{g_{00}}}.
\end{equation}
Integrating (\ref{arbitrarity in delta y}) with the above  $u^{\mu}$, we get 
\begin{equation}
m = \frac{F(x^{\rho})}{\sqrt{g_{00}(x^{\rho}, y)}},
\end{equation}
where $F$ is an arbitrary function of spacetime coordinates. We note that $m$ has to be invariant with respect to the transformation, 
\begin{eqnarray}
\bar{x}^{0} &=& \bar{x}^{0}(x^0, x^1, x^2, x^3),\nonumber \\
\bar{x}^{k} &=& \bar{x}^{k}(x^1, x^2, x^3),
\end{eqnarray}
 which leaves $u^{\mu}$ invariant. Therefore, $F$ is not a scalar function, but should transform as $\bar{F}(\bar{x}) = (\partial x^{0}/\partial \bar{x})F(x)$.

This is consistent with what we obtain from the definition of four-momentum $p_{\mu} = m u_{\mu}$. It implies $m = p_{\mu}u^{\mu}$, which for a particle at rest yields
\begin{equation}
\label{m from the definition of momentum}
m = \frac{p_{0}}{\sqrt{g_{00}}}.
\end{equation}
Thus, the function of integration $F$ is just $p_{0}$.  It should be noted that (\ref{arbitrarity in delta y}) and (\ref{spatial derivative of m}) link the derivatives $\partial g_{\mu\nu}/\partial y$ $\partial \Phi/\partial x^{\rho}$ to the variation of mass in the respective  directions, while $\partial g_{\mu\nu}/\partial x^{\rho}$ are related to the gravitational field.

\section{Formulae for  the rest mass}

In the original Randall $\&$ Sundrum scenario, only gravity is allowed to
propagate in the bulk, while all matter fields  are confined on the brane.
However,  the inclusion of matter
and gauge fields in the bulk has been extensively treated in
the literature (see \cite{Gustavo Burdman} and references therein).  
In particular, in models of Universal Extra Dimensions \cite{Appelquist} 
in
which {\it all} of the Standard Model fields are allowed to propagate in the bulk.

Therefore, for generality we do not restrict our discussion  to null geodesic motion in $5D$.
In this section we show how the observed rest mass in $4D$ is related to the metric and momentum along the extra dimension.  With this aim we multiply (\ref{geodesic equation in 5D}) by  
$g_{AD}$
\begin{equation}
g_{AD}\frac{d^2x^{A}}{dS^2} + g_{AD}\Gamma_{BC}^{A}U^{B}U^{C} = 0,
\end{equation}
and consider the equation for $D = 0$. After some manipulations we get 
\begin{equation}
\label{The D = 0 equation}
\frac{d}{f ds}\left(g_{\lambda 0}\frac{dx^{\lambda}}{f ds}\right)=  \frac{1}{2}\frac{\partial{g_{BC}}}{\partial x^{0}} U^{B}U^{C},
\end{equation}
where we have set
\begin{equation}
\label{definition of f}
dS = f ds, \;\;\mbox{with}\;\;\;f \equiv \sqrt{1 + \epsilon \Phi^2\left(\frac{dy}{ds}\right)^2}
\end{equation}

In order to interpret the term inside the
 bracket in (\ref{The D = 0 equation}), we introduce a 4D timelike unit vector 
field $\tau^{\mu}$, which is tangential to the time-coordinate
 $x^{0}$, viz.,
\begin{equation}
\tau^{\mu} = \frac{\delta^{\mu}_{0}}{\sqrt{g_{00}}}, \;\;\;\tau_{\mu} = \frac{g_{0\mu}}{\sqrt{g_{00}}}
\end{equation}
Also we introduce $\lambda_{\mu \nu}$, the projector onto the
 3-space, orthogonal to $\tau^{\mu}$
\begin{equation}
\label{space metric}
\lambda_{\mu \nu} \equiv \tau_{\mu} \tau_{\nu} - g_{\mu \nu}.
\end{equation}
Since $\lambda_{0j} = \lambda_{00} = 0$, the line element in $4D$
 becomes
\begin{equation}
\label{line element with proper time}
ds^{2} = g_{\mu\nu} dx^{\mu}x^{\nu} = d\tau^2 - dl^2,
\end{equation}
where $d\tau$ measures the proper time along an infinitesimal displacement $dx^{\mu}$ and $dl$ is the corresponding spatial length, viz.,
\begin{equation}
d\tau = \tau_{\mu}dx^{\mu}, \;\;\;dl = \sqrt{\lambda_{ij}dx^{i}dx^{j}}.
\end{equation}
Thus, 
\begin{equation}
\frac{d\tau}{ds} = \frac{1}{\sqrt{1 - v^2}},
\end{equation}
where $v^{2} \equiv \lambda_{ij} v^{i} v^{j}$ represents the square of 
the spatial three-velocity
\begin{equation}
\label{3 velocity}
v^{i} =   \frac{dx^{i}}{d\tau}.
\end{equation}
Thus we find
\begin{equation}
g_{0\mu}\frac{dx^{\mu}}{ds} = \frac{\sqrt{g_{00}}}{\sqrt{1 - v^2}}.
\end{equation}
Coming back to (\ref{The D = 0 equation}) we obtain
\begin{equation}
\label{The basic zero equation}
\frac{d}{fds}\left(\frac{\sqrt{g_{00}}}{f\sqrt{1 - v^2}}\right) = \frac{1}{2}\frac{\partial{g_{BC}}}{\partial x^{0}} U^{B}U^{C}. 
\end{equation}
In general relativity, the quantity inside the round bracket (with $f = 1$) is the energy $\cal{E}$ per unit mass
\begin{equation}
\label{energy per unit mass}
\frac{{\cal{E}}}{m} = \frac{\sqrt{g_{00}}}{\sqrt{1 - v^2}},
\end{equation}
which is constant when the gravitational field is independent of time. These equations suggest that the quantity 
\begin{equation}
\label{def. of mass for non-zero dS }
m = \frac{M}{f} = \frac{M}{\sqrt{1 + \epsilon \Phi^2 (dy/ds)^2}},
\end{equation}  
where $M$ is an arbitrary constant with the appropriate units, can be interpreted as the mass of a particle in $4D$. Notice that $m = M$ when  the motion is confined to hypersurfaces $y$ = constant.

In order to justify this interpretation, we have to show that our definition of mass satisfies (\ref{arbitrarity in delta y}) and (\ref{spatial derivative of m}) so we recover (\ref{Du/ds covariant}). From (\ref{def. of mass for non-zero dS }) we find
\begin{equation}
\label{dm/m}
\frac{dm}{m} = - \frac{df}{f}.
\end{equation}
From the definition of $f$ in (\ref{definition of f}) it follows that 
\begin{equation}
\label{differential of f}
f\frac{df}{ds} = \epsilon \Phi \frac{d \Phi}{ds} \left(\frac{dy}{ds}\right)^2 + \epsilon \Phi^2\frac{dy}{ds}\left(\frac{d^2 y}{ds^2}\right).
\end{equation}
Now, from (\ref{geodesic equation in 5D}) with $A = 4$ we obtain
\begin{equation}
\frac{d^2y}{ds^2} = \frac{1}{f}\left(\frac{df}{ds}\right)\frac{dy}{ds} - \Gamma^{4}_{AB}\frac{dx^A}{ds}\frac{dx^{B}}{ds}.
\end{equation}
Substituting into (\ref{differential of f}) we get
\begin{equation}
\label{(1/f)df/ds}
\frac{1}{f}\frac{df}{ds} = \epsilon \Phi \left(\frac{dy}{ds}\right)^2 - \epsilon \Phi^2\left(\frac{dy}{ds}\right)\Gamma^{4}_{AB}\frac{dx^A}{ds}\frac{dx^{B}}{ds}.
\end{equation}
After a simple calculation, and using (\ref{dm/m}) we find,
\begin{equation}
\label{dm/m in explicit form}
\frac{dm}{m} = \left[- \frac{1}{2}\frac{\partial g_{\mu\nu}}{\partial y}u^{\mu}u^{\nu}\right]\;dy + \left[\epsilon \Phi \frac{\partial \Phi}{\partial x^{\mu}}\left(\frac{dy}{ds}\right)^2 \right]dx^{\mu}. 
\end{equation}
Consequently, 
\begin{equation}
\label{partial derivatives of m}
\frac{1}{m}\frac{\partial m}{\partial y} = - \frac{1}{2}\frac{\partial g_{\mu\nu}}{\partial y}u^{\mu}u^{\nu}, \;\;\;\mbox{and}\;\;\;
\frac{1}{m}\frac{\partial m}{\partial x^{\mu}} = \epsilon \Phi \frac{\partial \Phi}{\partial x^{\mu}}\left(\frac{dy}{ds}\right)^2. 
\end{equation}
Thus, substituting these expressions into (\ref{equation of motion in terms of m}) we get exactly the equation of motion (\ref{Du/ds covariant}), obtained from the dimensional reduction of the geodesic equation in $5D$. In addition, we note that (\ref{def. of mass for non-zero dS }) is invariant under spacetime coordinate transformations ${\bar{x}}^{\mu} = {\bar{x}}^{\mu}(x^{\rho})$, which corresponds to the notion that $m = \sqrt{p_{\mu}p^{\mu}}$ is a scalar in $4D$.

\subsection{The rest mass in $4D$ for null geodesic motion in $5D$}

In the above discussion it is clear that $f \neq 0$, i.e., $dS \neq 0$. The question arises of how the motion along a null geodesic $(dS = 0)$ is observed in $4D$. The discussion is relevant to the original Randall $\&$ Sundrum  braneworld scenario  and other Kaluza-Klein models which postulate that the motion in $5D$ is along null geodesics \cite{Seahra4}. 
In this case the geodesic equation becomes 

 \begin{equation}
\label{geodesic equation in 5D for dS = 0}
\frac{d^2x^{A}}{d\lambda^2} + \Gamma_{BC}^{A}U^{B}U^{C} = 0, \;\;\;\;\mbox{with}\;\;\;U^{A} = \frac{dx^A}{d\lambda},
\end{equation}
where $\lambda$ is a parameter along the null geodesic in $5D$. Again, we can introduce a quantity $\bar{f}$ such that
\begin{equation}
\label{definition of fbar}
d \lambda = \bar{f} ds.
\end{equation}
It should be pointed out that, contrary to quantity $f$ defined in (\ref{definition of f}),  in general we have no  formulas relating $\bar{f}$ with other quantities in the theory, except in particular cases (see bellow in section $4.2$). 
However, following the same steps as above we would arrive at (\ref{The basic zero equation}) with $\bar{f}$ instead of $f$. So similarly, we can define
\begin{equation}
\label{definition of mass for dS = 0}
m = \frac{\bar{M}}{\bar{f}},
\end{equation}
where $\bar{M}$ is some constant with the appropriate units. 

Since $dS = \sqrt{1 + \epsilon \Phi^2(dy/ds)^2} = 0$, it follows that the extra dimension must be spacelike $\epsilon = - 1$, and    $ds = \Phi dy$. Consequently, $d\lambda = \bar{f}\Phi dy$ and 
\begin{equation}
\label{def. of mass for dS = 0}
m = \bar{M}\Phi\frac{dy}{d\lambda}.
\end{equation}
Taking the  differential of this quantity  with respect to $ds$ and using (\ref{geodesic equation in 5D for dS = 0}) with $A = 4$, we easily get
\begin{equation}
\frac{1}{m}\frac{dm}{ds} = - \frac{u^{\mu}u^{\nu}}{2\Phi}\frac{\partial g_{\mu\nu}}{\partial y} - \frac{u^{\mu}}{\Phi}\frac{\partial \Phi}{\partial x^{\mu}},
\end{equation}
which is formally obtained from (\ref{dm/m in explicit form}) by setting $(dy/ds) = 1/\Phi$ and $\epsilon = -1$. Clearly, this expression is equivalent to (\ref{partial derivatives of m}). Thus, when the mass defined through (\ref{def. of mass for dS = 0}) is substituted into (\ref{equation of motion in terms of m}) we obtain  the effective equations (\ref{Du/ds covariant}).

\section{Specific expressions for $m$}

In order to get a specific form for  $m$ one has to know $dy/ds$. Setting $A = 4$ in (\ref{geodesic equation in 5D}), after some manipulations we get\begin{equation}
\label{Geod. equation with A = 4}
\frac{d}{f ds}\left(\frac{\epsilon \Phi^2}{f}\frac{dy}{ds}\right) = \frac{1}{2}\frac{\partial g_{BC}}{\partial y}U^{B}U^{C}.
\end{equation} 

\subsection{Metric with no dependence on the extra coordinate}

In the case where the r.h.s. is zero, which in particular occurs when the metric is independent of $y$, we have
\begin{equation}
\frac{\Phi^2}{f}\frac{dy}{ds} = C,
\end{equation}
where $C$ is a constant. This is an important case because in the literature there are a huge number of solutions of the $5D$ equations with no dependence on the extra coordinate \cite{Wesson book}, \cite{LiuWessonPoncedeLeon}, \cite{Kokarev}. Thus,
\begin{equation}
\frac{dy}{ds} = \frac{C}{\Phi \sqrt{\Phi^2 - \epsilon C^2}}.
\end{equation}
Using this expression in (\ref{def. of mass for non-zero dS }) we find

\begin{equation}
\label{m for dS neqzero}
m = \frac{M\sqrt{\Phi^2 - \epsilon C^2}}{\Phi}.
\end{equation}
Clearly, $m = M$ for $dy/ds = 0$.
\subsection{Null geodesics in $5D$}
If $dS = 0$, we should replace $f \rightarrow \bar{f}$ in (\ref{Geod. equation with A = 4}). Thus,
\begin{equation}
\frac{\Phi^2}{\bar{f}}\frac{dy}{ds} = \bar{C},
\end{equation}
where $\bar{C}$ is a dimensionless constant. Consequently, $dy/d\lambda = \bar{C}/\Phi^2$. For $dS = 0$, it follows that $ds = \Phi dy$ (and $\epsilon = -1$) which gives $\bar{f} = \Phi/\bar{C}$, and therefore $d\lambda = \Phi ds/\bar{C}$.
 From (\ref{definition of mass for dS = 0}) we find
\begin{equation}
\label{definition of m for dS = 0}
m = \frac{\tilde{M}}{\Phi},
\end{equation}
where $\tilde{M} = \bar{M}\bar{C}$. Notice that both (\ref{m for dS neqzero}) and (\ref{definition of m for dS = 0}) satisfy (\ref{spatial derivative of m}).

When the metric is independent of the extra coordinate and $\Phi =$ constant, then $m$ is constant too. As a consequence, the r.h.s. of (\ref{equation of motion in terms of m}) is zero, meaning that embedding a $4D$ spacetime in a $5D$ manifold as 
\begin{equation}
dS^2  = g_{\mu\nu}(x^{\rho})dx^{\mu}dx^{\nu} \pm dy^2,
\end{equation}
produces no effects in $4D$.

\section{Conclusions}

In this paper we have used the principle of least action to obtain the equation of motion for a test particle in a four-dimensional spacetime embedded in a five-dimensional world with metric (\ref{general metric}).    

From our work emerges a clear physical picture. Specifically, that the deviation from the geodesic motion in $4D$ is due to the variation of the rest mass of a particle, which is induced by an explicit dependence of the spacetime metric on the extra coordinate. More explicitly, the rest mass (\ref{def. of mass for non-zero dS }) depends on $dy/ds$ which is governed by
\begin{equation}
\label{acceleration along y}
\frac{d^2y}{ds^2} = \frac{\epsilon u^{\mu}u^{\nu}}{2 \Phi^2}\frac{\partial g_{\mu \nu}}{\partial y} + \frac{dy}{ds}\left[\frac{1}{f}\frac{df}{ds} + \frac{2 u^{\mu}}{\Phi}\frac{\partial \Phi}{\partial x^{\mu}} + \frac{1}{\Phi}\frac{\partial \Phi}{\partial y}\frac{dy}{ds}\right],
\end{equation}
where $(df/fds)$ is given by (\ref{(1/f)df/ds}). This equation indicates that, even if at some initial moment $(dy/ds) = 0$, the non-trivial dependence of the metric on the extra variable implies $(dy/ds) \neq 0$ the next moment, which will be perceived by an observer in $4D$ as a variation in the rest mass of a particle.

Notice that the scalar field $\Phi$ by itself does not generate momentum along the extra dimension. Indeed, in the case where the metric is independent of the extra coordinate, if   $(dy/ds) = 0$ at some moment, then $d^2y/ds^2 = 0$. Which means that $dy/ds$ will continue to be zero along the motion. Consequently, the geodesic motion is on a hypersurface $y =$ constant, which in $4D$ will be perceived as a particle of constant mass $m = M$. 
However, one would expect that any small perturbation along $y$ would be enhanced by the scalar field,  building up momentum  along the extra dimension, which in $4D$ will be perceived  as a variation of the rest mass. Clearly, the same reasoning holds for  the case of null geodesics in $5D$ where $m$ is given by (\ref{definition of mass for dS = 0}). 

The question may arise about the connection between the four momentum and $m$. In order to see this, let us construct the five-dimensional quantity
\begin{equation}
\label{momentum in 5D}
P^A = M \frac{dx^A}{dS}.
\end{equation}
The square of this quantity is $M^2$. Namely,
\begin{equation}
P_{A}P^{A} = g_{\mu\nu}P^{\mu}P^{\nu} + \epsilon \Phi^2 \frac{dy}{dS} = M^2.
\end{equation}
Using (\ref{definition of f}) and rearranging terms we obtain
\begin{equation}
g_{\mu\nu}P^{\mu}P^{\nu} = m^2,
\end{equation}
where $m$ is given by (\ref{def. of mass for non-zero dS }). This suggests that the spacetime part of $P^{A}$, which coincides with the projection onto $4D$, should  be  identified with the four-momentum $p^{\mu}$, viz.,
\begin{equation}
p^{\mu} = P^{\mu} = M\frac{dx^{\mu}}{dS} = \left(\frac{M}{f}\right)u^{\mu} = mu^{\mu},
\end{equation}
in agreement with the usual definition of $4$-momentum in relativity.

Some authors postulate that motion in $5D$ is along null geodesics, similarly to braneworld theory. In this case, the extra dimension must be spacelike and the motion along null geodesics is  observed in $4D$ as particles with an effective mass, which is given by (\ref{definition of mass for dS = 0}) and (\ref{def. of mass for dS = 0}). 
 In this case, we introduce the quantity
\begin{equation}
\label{P for null geodesics}
 {\bar{P}}^A = \bar{M}\frac{dx^A}{d \lambda},
\end{equation}
 where $\bar{M}$ is a constant introduced for dimensional consistency, and $\lambda$ is the affine parameter along the null geodesic introduced  in (\ref{geodesic equation in 5D for dS = 0}). Clearly, in this case ${\bar{P}}_{A}{\bar{P}}^{A} = 0$. Thus, using (\ref{definition of fbar}) we obtain
$(\epsilon = - 1)$
\begin{equation}
g_{\mu\nu}{\bar{P}}^{\mu}{\bar{P}}^{\nu} = \frac{\bar{M}\Phi^2}{{\bar{f}}^2}\left(\frac{dy}{ds}\right)^2.
\end{equation}
For null geodesics $(dy/ds) = 1/\Phi$. Thus, 

\begin{equation}
g_{\mu\nu}{\bar{P}}^{\mu}{\bar{P}}^{\nu} = m^2, \;\;\;\;p^{\mu} = {\bar{P}}^{\mu} = m u^{\mu}, \;\;\;\mbox{with}\;\;\;m = \bar{M}\Phi \frac{dy}{d\lambda},
\end{equation} 
as expected. A particular example is provided by (\ref{definition of m for dS = 0}).

For completeness let us emphasize the main differences and similarities between  null and non-null geodesic motion in $5D$, as observed in $4D$: (i) For null geodesics in $5D$, the observed mass in $4D$ is a consequence of non-zero momentum  along the extra dimension;  it  is massless  if the motion is confined to a hypersurface $y = $constant; (ii) For non-null geodesics in $5D$,  the observed particle in $4D$ is a massive one, even in the absence of momentum along $y$; (iii) If $dy/ds =$ constant $\neq 0$ $(dy/d\lambda =$ constant $\neq 0)$, then the  variation of rest mass  is a consequence of the scalar field $\Phi$; (iv)  If $\Phi =$ constant, then the variation of mass is a consequence of the dependence of the $4D$ metric on the extra coordinate.


\begin{thebibliography}{99}
\bibitem{Arkani1}{N. Arkani-Hamed, S. Dimipoulos and G. Dvali, {\em Phys. Lett.} {\bf B 429}, 263(1998), hep-ph/9803315.}
\bibitem{Arkani2}{N. Arkani-Hamed, S. Dimipoulos and G. Dvali, {\em Phys. Rev.} {\bf D 59}, 086004(1999), hep-ph/9807344.}
\bibitem{Arkani3}{I. Antoniadis, N. Arkani-Hamed, S. Dimipoulos and G. Dvali, {\em Phys. Lett.} {\bf B 436}, 257(1998), hep-ph/9804398.}
\bibitem{RS1}{L. Randall and R. Sundrum, {\em Mod. Phys. Lett.} {\bf A13}, 2807(1998), hep-ph/9905221.}
\bibitem{RS2}{L. Randall and R. Sundrum, {\em Phys. Rev. Lett.} {\bf 83}, 4690(1999), hp-th/9906064.}
\bibitem{JPdeL 1}{J. Ponce de Leon, {\em Gen. Rel. Grav.} {\bf 20}, 539(1988).}
\bibitem{WessonJPdeL}{P.S. Wesson and J. Ponce de Leon, {\em J. Math. Phys.} {bf 33}, 3883(1992).}
\bibitem{JPdeL Wesson}{J. Ponce de Leon and P.S. Wesson, {\em J. Math. Phys.} {\bf 34}, 4080(1993).}
\bibitem{Wesson book}{P.S. Wesson, {\em Space-Time-Matter} (World Scientific Publishing Co. Pte. Ltd. 1999).}
\bibitem{EMT}{J. Ponce de Leon, {\em Int.J.Mod.Phys.} {\bf D11}, 1355(2002), gr-qc/0105120.}
\bibitem{Campbell}{J.E. Campbell,  {\em A Course of Differential Goemetry} (Clarendon, Oxford, 1926).}
\bibitem{Rippl}{S. Rippl, C. Romero and R. Tavakol, {\em Class. Quant. Grav.} {\bf 12}, 2411(1995), gr-qc/9511016.}
\bibitem{Romero1}{C. Romero, R. Tavakol and  R. Zalaletdinov, {\em Gen. Rel. Grav.} {\bf 28}, 365 (1996).}
\bibitem{Lidsey}{J.E. Lidsey, C. Romero, R. Tavakol and S. Rippl, {\em Class. Quant. Grav.} {\bf 14}, 865(1997), gr-qc/9907040.}
\bibitem{MashhoonWesson}{B. Mashhoon, P.S. Wesson, and H. Liu, {\em Gen. Rel. Grav.} {\bf 30}, 555(1998).}
\bibitem{WessonMashhoon}{P.S. Wesson, B. Mashhoon, H. Liu, W.N. Sajko, {\em Phys. Lett.} {\bf B456}, 34(1999).}
\bibitem{Youm1}{D. Youm, {\em Phys. Rev.} {\bf D62}, 084002 (2000), hep-th/0004144.}
\bibitem{Youm2}{D. Youm, {\em Mod. Phys. Lett.} {\bf A16}, 2371(2001), hep-th/0110013.}
\bibitem{JPdeLforce}{J. Ponce de Leon, {Phys. Lett.} {\bf B523}, 311(2001), gr-qc/0110063.}
\bibitem{Seahra3}{S.S. Seahra, {\em Phys. Rev.} {\bf D 65}, 124004(2002), gr-qc/0204032.}
\bibitem{Seahra4}{S.S. Seahra, and P.S. Wesson, {\em Gen. Rel. Grav.} {\bf 33}, 1731(2001), gr-qc/0105041.}
\bibitem{MashhoonLiu}{B. Mashhoon, H. Liu, and P.S. Wesson, {\em Phys. Lett.} {\bf B 331}, 305(1994).}
\bibitem{LiuMashhoon}{H. Liu, and B. Mashhoon, {\em Phys. Lett.} {\bf A 272}, 26(2000), gr-qc/0005079.}
\bibitem{Billyard}{A.P. Billyard and W.N. Sajko, {\em Gen. Rel. Grav.} {\bf 33}, 1929(2001), gr-qc/0105074.}
\bibitem{DynKK}{J. Ponce de Leon, {\em Int.J.Mod.Phys.}  {\bf D12},  757(2003), gr-qc/0209013.}
\bibitem{Romero2}{F. Dahia, E. M. Monte and C. Romero, {\em Mod.Phys.Lett.}  {\bf A18},  1773(2003), gr-qc/0303044.}
\bibitem{Landau and Lifshitz}{L. Landau and E. Lifshitz, {\em The Classical Theory of Fields} (Pergamon, New York, 1975), 4th ed.}
\bibitem{comment}{The action is denoted by $S$ in Landau and Lifshits, here we use $I$ to avoid confusions with the interval in $5D$. A comment found as a footnote on page $25$ of reference  (27) might be  of relevance here. ``Strictly speaking, the principle of least action asserts that the integral $S$ must be a minimum only for infinitesimal lengths of the path of integration. For paths of arbitrary length we can say only that $S$ must be an extremum, not necessarily a minimum."}
\bibitem{Gustavo Burdman}{G. Burdman, {\em Physical Review} {\bf D66}, 076003(2002), hep-ph/0205329.}
\bibitem{Appelquist}{T. Appelquist, Hsin-Chia Cheng, B.A. Dobrescu, {\em Phys.Rev.} {\bf D64},  035002(2001), 
hep-ph/0012100.}
\bibitem{LiuWessonPoncedeLeon}{H. Liu, P.S. Wesson and J. Ponce de Leon, {\em J. Math. Phys.} {\bf 34}, 4070(1993).}
\bibitem{Kokarev}{S. Kokarev and V.G.Kretchen, {\em Grav. Cosmol.} {\bf 2}, 99(1996).}
\end{thebibliography}
\end{document}